\documentclass[submission,copyright,creativecommons]{eptcs}
\providecommand{\event}{ICLP 2024} 

\usepackage{iftex}

\ifpdf
  \usepackage{underscore}         
  \usepackage[T1]{fontenc}        
\else
  \usepackage{breakurl}           
\fi

\title{Architecture for Simulating Behavior Mode Changes in Norm-Aware Autonomous Agents}
\author{Sean Glaze
\institute{Miami University\\
Ohio, USA}
\email{glazesc@miamioh.edu}
\and
Daniela Inclezan
\institute{Miami University\\
Ohio, USA}
\email{inclezd@miamioh.edu}
}
\def\titlerunning{Architecture for Simulating Behavior Mode Changes}
\def\authorrunning{S. Glaze \& D. Inclezan}

\usepackage{amsmath}
\usepackage{graphicx}
\usepackage{multirow}
\usepackage{xspace}
\usepackage{comment}
\usepackage{hyperref}
\usepackage{pdfcomment}
\usepackage{float}
\usepackage[mathscr]{euscript}
\usepackage{xcolor}
\usepackage{colortbl}

\usepackage{caption}

\def\aopl{$\mathscr{AOPL}$\xspace}
\def\apia{$\mathscr{APIA}$\xspace}
\def\ald{$\mathscr{AL}_d$\xspace}

\def\p{$\mathscr{P}$\xspace}
\def\t{$\mathscr{T}$\xspace}
\def\lp{$lp(\mathscr{P}, \sigma)$\xspace}
\def\Ps{$\mathscr{P}(\sigma)$\xspace}
\def\ev{$\langle \sigma, a \rangle$\xspace}

\newtheorem{definition}{Definition}

\begin{document}
\maketitle

\begin{abstract}
This paper presents an architecture for simulating the actions 
of a norm-aware intelligent agent whose behavior with respect to norm compliance is set, and can later be changed, by a human controller. Updating an agent's behavior mode from a norm-abiding to a riskier one may be relevant when the agent is involved in time-sensitive rescue operations, for example. We base our work on the \textit{Authorization and Obligation Policy Language} \aopl designed by Gelfond and Lobo for the specification of norms. We introduce an architecture and a prototype software system that can be used to simulate an agent's plans under different behavior modes that can later be changed by the controller. We envision such software to be useful to policy makers, as they can more readily understand how agents may act in certain situations based on the agents' attitudes towards norm-compliance. Policy makers may then refine their policies if simulations show unwanted consequences. 
\end{abstract}



\section{Introduction}

This paper introduces an architecture for simulating the actions to be taken by an intelligent agent that is aware of norms (i.e., policies\footnote{We use the words \textit{norms} and \textit{policy} interchangeably in this paper.}) governing the domain in which it acts. We assume that different agents may exhibit different behavior modes with respect to norm-compliance: some may be very cautious and norm-abiding, while others may exhibit a riskier behavior. We consider the case in which the behavior mode under which an agent operates is set by a human controller who can update it if needed, for instance in cases when the agent is involved in a time-sensitive rescue operation. 

This architecture is relevant to modeling physical intelligent agents that act autonomously, for instance robots deployed in harsh environments (underwater, on Mars, in mines), and whose settings may be re-adjusted by a human controller if the circumstances require it, but this is done sparingly in emergency situations. The architecture is also crucial 
to simulating the behavior of human agents with different norm-abiding attitudes, especially if such attitudes change over time. This can be of value to policy makers as testing their policies on different human agent models can lead to policy improvement, if unwanted consequences are observed in the simulation (similarly to work by Corapi et al. \cite{CorapiVPRS10} on creating \textit{use cases} for policy development and refinement).

In our work, we utilize the \textit{Authorization and Obligation Policy Language} (\aopl) by Gelfond and Lobo \cite{gl08} for norm specification, due to its close connection to Answer Set Programming (ASP). In fact, the semantics of \aopl and the notion of norm-compliance are defined via a translation into ASP. This allows us to leverage existing ASP methodologies for representing dynamic domains, planning, and creating agent architectures, as well as ASP solvers like \textsc{Clingo} (\url{https://potassco.org/clingo/}) 
or DLV (\url{https://www.dlvsystem.it/dlvsite/}). 
A reason for using \aopl instead of representing norms directly in ASP (as soft constraints for example) is that \aopl provides policy analysis capabilities \cite{di23}, which are important for checking that a policy imposed on an agent is actually valid and unambiguous; similar policy analysis would be difficult to conduct without the use of a higher-level language for norm specification. In addition to \aopl, we build upon work on norm-aware autonomous agents by \cite{hi23} who introduced the notion of behavior modes with respect to norm-compliance. 

Our main contributions are two-fold: (1) we \textbf{introduce an architecture} for norm-aware autonomous agents who may exhibit different behavior modes and may experience changes between bahavior modes; and (2) we \textbf{implement a software system} for the simulation of an agent's actions under behavior modes that may change over time.

In the remainder of the paper, we start with background information in Section \ref{sec:background} and provide a motivating example in Section \ref{sec:example}. We introduce our architecture in Section \ref{sec:architecture}, then our software system for simulation in Section \ref{sec:gui}, and examine the system's evaluation in Section \ref{sec:evaluation}. We discuss related work in Section \ref{sec:related} and end with conclusions in Section \ref{sec:conclusions}.


\section{Background}
\label{sec:background}
In this section we introduce the norm-specification language \aopl and behavior modes for norm-aware agents. We assume that readers are familiar with ASP and otherwise direct them to outside resources on ASP \cite{gl91,mt99,gk14}. 

\subsection{Norm-Specification Language \aopl}
\label{sec:aopl}
Gelfond and Lobo \cite{gl08} 
introduced the \textit{Authorization and Obligation Policy Language} (\aopl) for specifying policies for an intelligent agent acting in a dynamic environment. A policy is a collection of authorization and obligation statements. 
An {\em authorization} indicates whether an agent's action is permitted or not, and under which conditions. 
An {\em obligation} describes whether an agent is obligated or not obligated to perform a specific action under certain conditions. 
An \aopl policy works in conjunction with a dynamic system description of the agent's environment written in an action language such as \ald \cite{gi13}. 
The signature of the dynamic system description includes predicates denoting {\em sorts} for the elements in the domain;
{\em fluents} (i.e., properties of the domain that may be changed by actions); and {\em actions}.
An \ald system description defines the domain's transition diagram 
whose states are complete and consistent sets of fluent literals and whose arcs are labeled by action atoms (shortly {\em actions}). 

{\bf The signature of an \aopl policy} includes the signature of the associated dynamic system and additional predicates $permitted$ for authorizations,
$obl$ for obligations, and \textit{prefer} for specifying preferences between authorizations or obligations. A \textit{prefer} atom is created from the predicate \textit{prefer}; similarly for $permitted$ and $obl$ atoms.

An \aopl \textbf{policy} \p is a finite collection of statements of the form:
\begin{subequations} \label{eq2}
    \begin{align}
            & \ \ permitted\left(e\right) & \textbf{ if } \ cond \label{eq1_1}\\[-0.3em]
            & \neg permitted\left(e\right) & \textbf{ if } \ cond \label{eq1_2}\\[-0.3em]
		& \ \ obl\left(h\right) & \textbf{ if } \ cond \label{eq1_3}\\[-0.3em]
            & \neg obl\left(h\right) & \textbf{ if } \ cond\label{eq1_4}\\[-0.3em]
        d: \textbf{normally } & \ \ permitted(e) & \textbf{ if } \ cond \label{eq2_1}\\[-0.3em]
        d: \textbf{normally } & \neg permitted(e) & \textbf{ if } \ cond \label{eq2_2}\\[-0.3em]
        d: \textbf{normally } & \ \ obl(h) & \textbf{ if } \ cond \label{eq2_3}\\[-0.3em]
        d: \textbf{normally } & \neg obl(h) & \textbf{ if } 
        \ cond \label{eq2_4}\\[-0.3em]
	    	& \ \ \textit{prefer}(d_i, d_j) & \label{eq2_5}
    \end{align}
\end{subequations}
where $e$ is an elementary action; $h$ is a happening (i.e., an elementary action or its negation\footnote{If $obl(\neg e)$ is true, 
then the agent must not execute $e$.}); 
$cond$ is a 
set of atoms of the signature, except for atoms containing the predicate \textit{prefer}; $d$ in (\ref{eq2_1})-(\ref{eq2_4}) and $d_i$, $d_j$ in (\ref{eq2_5}) denote defeasible rule labels.
Rules (\ref{eq1_1})-(\ref{eq1_4}) encode {\em strict} policy statements, while rules (\ref{eq2_1})-(\ref{eq2_4}) encode {\em defeasible} statements. 
Rule (\ref{eq2_5}) captures {\em priorities} between defeasible statements.

The\textbf{ semantics} of an \aopl policy determine a mapping \Ps from states of a transition diagram \t into a collection of $permitted$ and $obl$ literals. To formally describe the semantics of \aopl, a translation of a policy ${\cal P}$ and a state $\sigma$ of the transition diagram into ASP is defined as $lp({\cal P}, \sigma)$ 
as described in the paper by Gelfond and Lobo \cite{gl08}.
Properties of an \aopl policy \p are defined in terms of the answer sets of the logic program $lp({\cal P}, \sigma)$ expanded with appropriate rules.

The following definitions by Gelfond and Lobo are relevant to our work (original definition numbers in parenthesis). In what follows $a$ denotes a (possibly) compound action (i.e., a set of simultaneously executed elementary actions), while $e$ refers to an elementary action. An event \ev is a pair consisting of a state $\sigma$ and an 
action $a$ executed in $\sigma$.

\begin{definition}[Consistency and Categoricity -- Defs. 3 and 6]
\label{def:consistency}
A policy \p for \t is called {\em consistent} if for every state $\sigma$ of \t, the logic program \lp has an answer set. It is called {\em categorical} if \lp has \textit{exactly} one answer set.
\end{definition}

\vspace{-5pt}
\begin{definition}[Policy Compliance for Authorizations and Obligations -- Defs. 4, 5, and 9]
\label{def:auth_compliance}

$\bullet\ $ An event \ev is {\em strongly-compliant} with authorization policy \p if for every $e \in a$ 
the logic program \lp entails $permitted(e)$.

\noindent
$\bullet\ $ An event \ev is {\em weakly-compliant} with authorization policy \p if for every $e \in a$ 
the logic program \lp does not entail $\neg permitted(e)$.

\noindent
$\bullet\ $ An event \ev is {\em non-compliant} with authorization policy \p if for every $e \in a$ 
the logic program \lp entails $\neg permitted(e)$.

\noindent
$\bullet\ $ An event \ev is {\em compliant} with obligation policy \p if 

\noindent
$\ \ - $ For every $obl(e) \in$ \Ps we have that $e \in a$, and

\noindent
$\ \ - $ For every $obl(\neg e) \in$ \Ps we have that $e \notin a$.

\end{definition}

\subsection{Behavior Modes in Norm-Aware Autonomous Agents}
\label{sec:charlie}

Harders and Inclezan \cite{hi23} introduced an ASP framework for plan selection for norm-aware autonomous agents, where norms were specified in \aopl. They built upon observations by Inclezan \cite{di23} indicating that, for categorical \aopl policies, all strongly-compliant actions are also weakly-compliant w.r.t. authorizations and that modality conflicts between authorizations and obligations may occur when the \aopl policy simultaneously contains obligations and prohibitions to execute an action. 
Instead, the notion of an \textit{underspecified} event was introduced to denote an event that is not explicitly known to be compliant nor non-compliant w.r.t. authorizations, and a \textit{modality ambiguous} event as an event arising from a modality conflict. Harders and Inclezan proposed that agents may have different attitudes towards norm compliance that would impact the selection of the ``best" plan. They called these attitudes \textit{behavior modes} and introduced different metrics that can be used to express them. They also presented some predefined agent behavior modes, defined as follows:
(a) \textbf{Safe Behavior Mode} -- prioritizes events that are explicitly known to be compliant and does not execute non-compliant actions;
(b) \textbf{Normal Behavior Mode} -- prioritizes plan length and then actions explicitly known to be compliant, while not executing non-compliant actions; and 
(c) \textbf{Risky Behavior Mode} -- disregards policy rules, but does not go out of its way to break rules either. 

\section{Example}
\label{sec:example}

For illustration purposes, consider a \emph{Mining Domain} consisting of a 3x3 square grid of locations with an associated risk level (low, medium, or high) and three ores (gold, silver, and iron) with unique locations across the grid. The mining robot can collect ores or move between adjacent locations. The mining robot's goal is to collect all three ores. The norm that is imposed in this domain is that the collection of ores must happen in the sequence: gold first, then silver, and finally iron. 

The mining robot has three \emph{behavior modes}: Safe, Normal, and Risky, as defined in Section \ref{sec:charlie}, but expanded with some additional policies. The Safe agent is obligated to move only through low-risk locations, the Normal agent is obligated to only move through low or medium-risk locations, and the Risky agent moves freely throughout the grid with no regard for the risk level of locations. Furthermore, as the Risky behavior mode does not have any regard for policies, an agent in this mode will collect ores in whichever order leads to the shortest plan possible. 

We will now discuss a specific scenario within the mining domain shown in Figure 1. 
In this illustration, locations are labeled l0 to l8, with connected locations indicated by a black line. Each location is colored green, yellow, or red to indicate a low, medium, or high-risk level, respectively. The mining robot is depicted in its initial location and the locations of ores are indicated by their corresponding labels in the periodic table. 

\medskip
\begin{minipage}{\textwidth}
  \begin{minipage}
  {0.43\textwidth}
    \centering
    \includegraphics[width=0.9\textwidth]
    {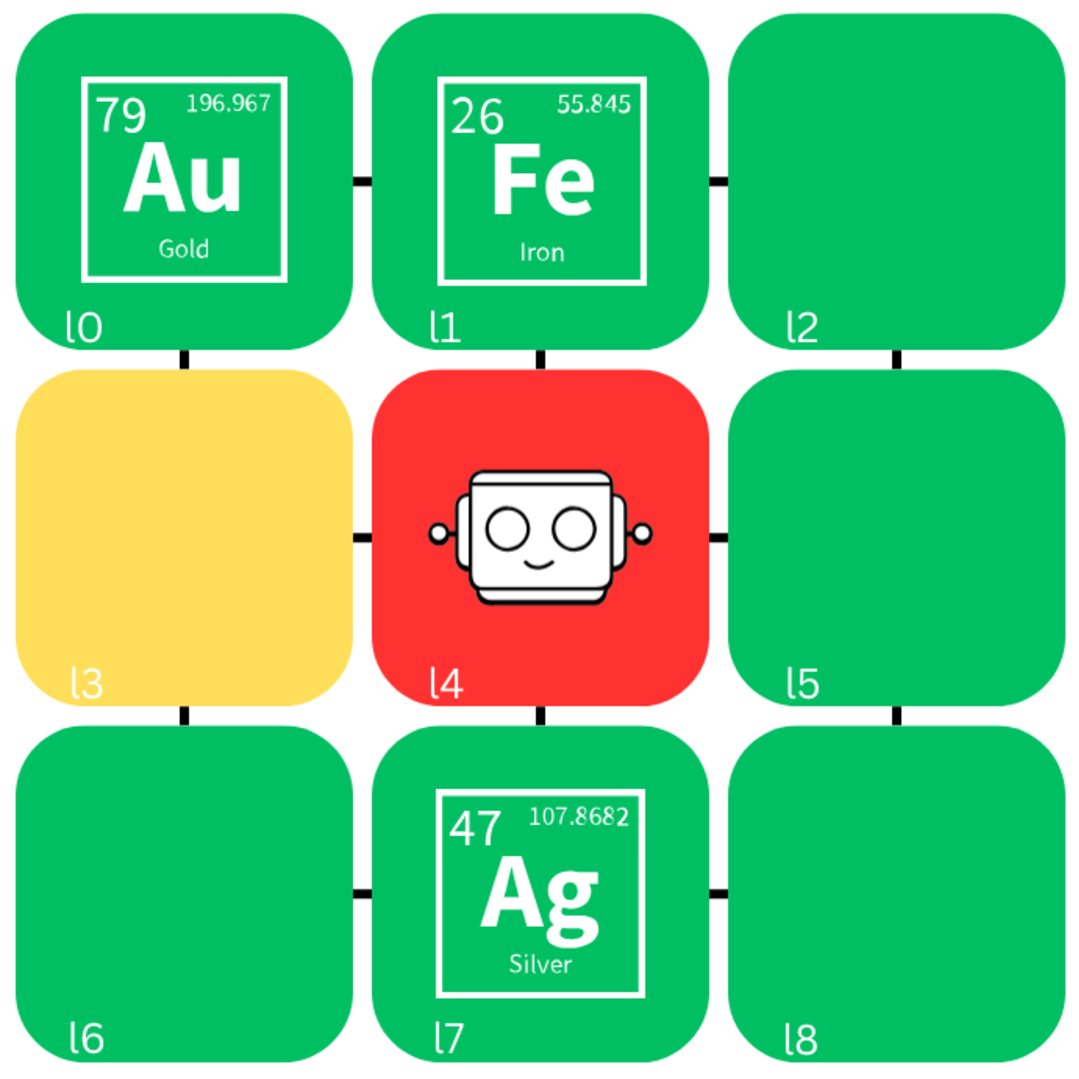}
    \label{fig:MiningRobotScenario}
    \captionof{figure}{Mining Domain: Sample Scenario}
  \end{minipage}
  \hfill
  \begin{minipage}
  {0.57\textwidth}
    \centering
    \small
    \captionof{table}{Plan with Behavior Mode Changes}
    \label{tbl:bm_changes}
    \begin{tabular}{l} 
    \hline\hline
    *** Begin in Safe Mode ***\\
    \hline
    \verb|0. Move from l4 to l1|\\
    \verb|1. Move from l1 to l0|\\
    \verb|2. Collect gold|\\
    \hline
    *** Change to Normal Mode ***\\
    \hline
    \verb|3. Move from l0 to l3|\\
    \verb|4. Move from l3 to l6|\\
    \verb|5. Move from l6 to l7|\\
    \verb|6. Collect silver|\\
    \hline
    *** Change to Risky Mode ***\\
    \hline
    \verb|7. Move from l7 to l4|\\
    \verb|8. Move from l4 to l1|\\
    \verb|9. Collect iron|\\
    \hline\hline
    
   \end{tabular}
 \end{minipage}
\end{minipage}
\normalsize

\medskip
Table 2 shows the plans that the agent devises depending on its behavior mode. 
This scenario illustrates general outcomes, where cautious behavior modes result in longer plans, while the Risky behavior mode generates the shortest plans, with a trade-off of a detrimental effect on safety and policy-compliance. The longer plan devised in the Safe mode is caused by the inability to move through location l3, which is not a low-risk location. The Risky mining robot produces the shortest plan because it disregards location risk and the policy of collecting ores in a specific order.

Now let's consider behavior mode changes in this scenario. Table 1 
shows the plan that
is generated by a mining robot that begins in Safe mode, is switched by the controller to Normal mode at time step 3, and is switched again to Risky mode at time step 7. In case of emergency or changing priorities, a more risky behavior may be desired by the controller of such a robot, even though it does result in a higher degree of danger for the robot. We want our architecture to simulate such behavior mode modifications.

\begin{table}[h!tb]
\centering
\caption{Plans for the Scenario in Fig. 1 for Different Behavior Modes}
\label{tbl:motivating_example}
    \begin{tabular}{lll}
      \hline\hline
Safe   Behavior Mode    & Normal Behavior Mode   & Risky Behavior Mode   \\
\hline
$\ $ 0. Move from l4 to l1 $\ \ \ \ \ \ \ \ $      & $\ $ 0. Move from l4 to l1   $\ \ \ \ \ \ \ \ $ & 0. Move from l4 to l7 \\
$\ $ 1. Move from l1 to l0      & $\ $ 1. Move from l1 to l0    & 1. Collect silver     \\
$\ $ 2. Collect gold            & $\ $ 2. Collect gold          & 2. Move from l7 to l4\\
$\ $ 3. Move from l0 to l1      & $\ $ 3. Move from l0 to l3    & 3. Move from l4 to l1 \\
$\ $ 4. Move from l1 to l2      & $\ $ 4. Move from l3 to l6    & 4. Collect iron       \\
$\ $ 5. Move from l2 to l5      & $\ $ 5. Move from l6 to l7    & 5. Move from l1 to l0 \\
$\ $ 6. Move from l5 to l8      & $\ $ 6. Collect silver        & 6. Collect gold      \\
$\ $ 7. Move from l8 to l7      & $\ $ 7. Move from l7 to l6    &                       \\
$\ $ 8. Collect silver          & $\ $ 8. Move from l6 to l3    &                       \\
$\ $ 9. Move from l7 to l8      & $\ $ 9. Move from l3 to l0    &                       \\
10. Move from l8 to l5          & 10. Move from l0 to l1   &                       \\
11. Move from l5 to l2          & 11. Collect iron         &                       \\
12. Move from l2 to l1          &                        &                       \\
13. Collect iron                &                        &   \\                   
      \hline\hline
    \end{tabular}
\end{table}

\section{Architecture}
\label{sec:architecture}
We identified three questions that needed to be answered during the development of this architecture, outlined below together with our design decisions:

\begin{itemize}
    \item \textit{How will an agent adjust its plan when its behavior mode is modified?}

    \textbf{Design Decision:} The agent will devise a new plan with its new behavior mode, starting at the time step that the behavior mode modification is set to take effect.
    
    \item \textit{How will the agent's memory mechanism work with respect to already executed actions of a plan? In other words, how will the agent deal with prior actions that may not satisfy the definition of its new behavior mode?}

    \textbf{Design Decision:} The agent remembers the behavior mode under which it operated at each point in time and checks requirement satisfaction w.r.t. to the behavior mode settings in place when the action was executed, to mimic real world situations where new laws are not applied retrospectively.
    
    \item \textit{Does the agent need to be aware that its behavior mode is liable to be modified at later points in time?}

    \textbf{Design Decision:} The agent is not explicitly aware that its behavior mode can be modified. However, we introduce the concept of \emph{subgoals} so that the agent can strive to partially complete its overall goal if its current behavior mode prevent completing the goal as a whole. 
\end{itemize}


The proposed architecture consists of two distinct components that work in conjunction: an ASP Component and a Python Component, discussed in detail in the following subsections. Figure \ref{fig:ArchitectureIllustration} provides an overall view of the proposed architecture.

\begin{minipage}{\textwidth}
    \centering
    \includegraphics[width=0.7\textwidth]{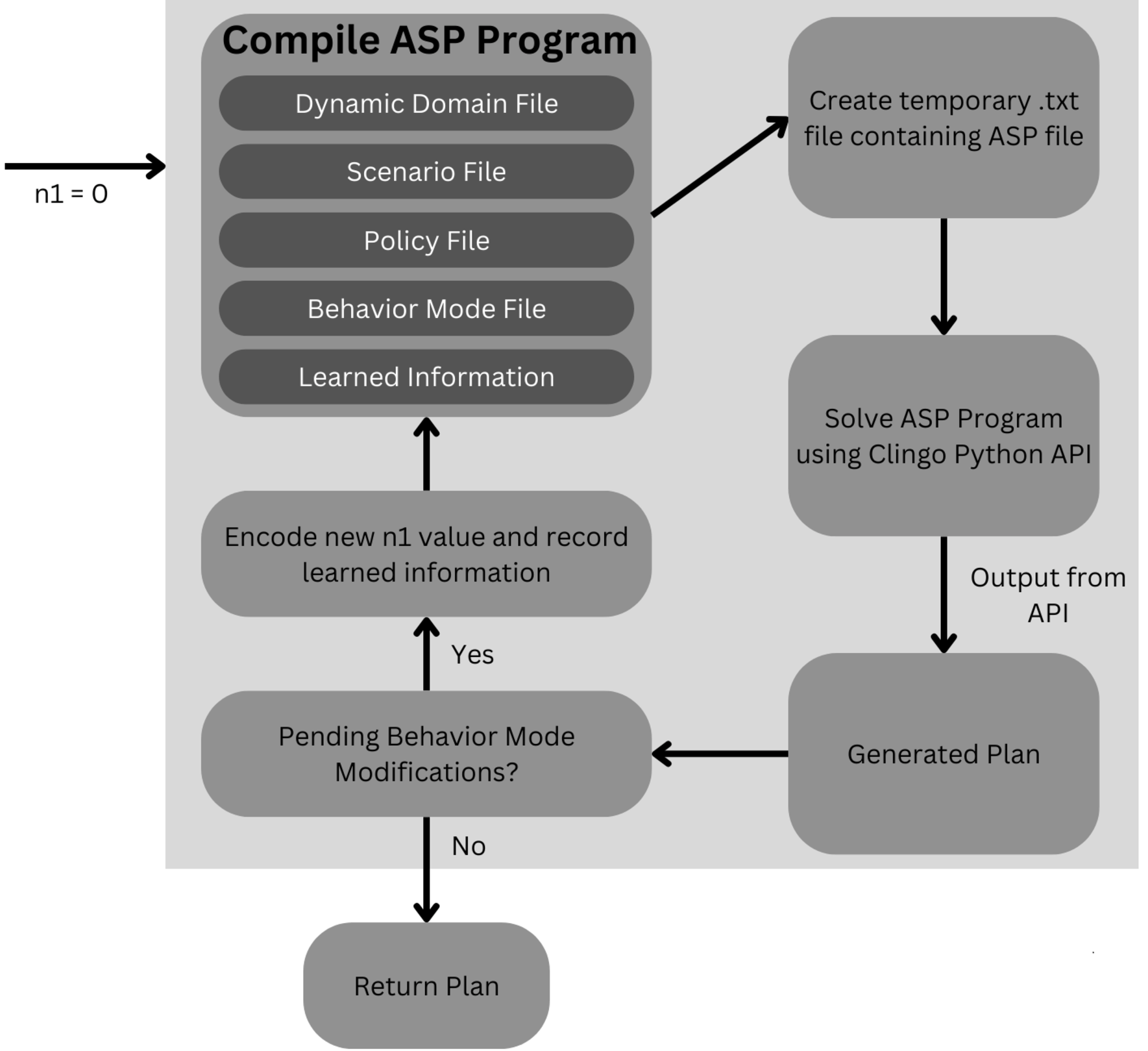}
    \label{fig:ArchitectureIllustration}
    \captionof{figure}{Illustration of Proposed Architecture}
\end{minipage}

\subsection{ASP Component}

The ASP Component consists of five pieces: the dynamic domain encoding, scenario encoding, policy encoding, behavior mode encoding, and learned information.

The \textit{dynamic domain encoding} contains the objects, statics, fluents, actions, and axioms that define the dynamic domain, encoded according to established ASP methodologies \cite{gk14}. In the Mining Domain, the objects are $locations$ and $ores$. Statics are used to describe whether two locations are connected, as well as the risk level of a location ($has\_risk\_level(l, level)$). Inertial fluents at play are the agent's location ($at\_loc(l)$); whether the agent possesses a certain ore or not ($has\_ore(o)$); and the locations of the ores ($ore\_loc(o, l)$). The agent can perform two actions: move from one location to another ($move(l_1, l_2)$); and collect an ore ($collect(o)$). It has an additional action $wait$ that does not change the state of the domain.

The \textit{scenario encoding} consists of a list of facts that are true at time step 0. For the Mining Domain, this means specifying the risk level of each of the locations, the initial location of the agent, and the location of the ores. 

The \textit{policy encoding} contains the ASP translations of the \aopl policies that govern the dynamic domain. For the Mining Domain, there is only one policy that applies by default: the agent is obligated to collect the ores in the sequence gold, silver, iron. This is encoded in two separate \aopl rules -- one that says the agent is obligated not to collect silver unless it possesses gold and another that says the agent is obligated not to collect iron unless it possesses silver:
$$
\begin{array}{lll}
    obl(\neg collect(silver)) & \textbf{if} & \neg has\_ore(gold)\\
    obl(\neg collect(iron)) & \textbf{if} & \neg has\_ore(silver)\\
\end{array}
$$

The \textit{behavior mode encoding} in this architecture considers tree behavior modes: Safe, Normal, and Risky. As previously mentioned, the exact definitions of these behavior modes are meant to be tailored to the dynamic domain's specific needs. For example, in the Mining Domain, the Safe and Normal agents are under additional policies. Specifically, the Safe agent is obligated not to move through high or medium-risk locations, and the Normal agent is obligated not to move through high-risk areas. These rules are written in \aopl, as shown below for the Safe agent, and are translated into ASP to be used in our architecture: 
$$
\begin{array}{lll}
    obl(\neg move(L1, L2)) & \textbf{if} & has\_risk\_level(L_2, high)\\
    obl(\neg move(L1, L2)) & \textbf{if} & has\_risk\_level(L_2, medium)\\
\end{array}
$$

\noindent
The \textit{behavior mode encoding} also contains general ASP rule for planning, such as: 
\small
\begin{verbatim}
    1 {occurs(A,I) : action(A)} 1:- step(I), I >= n1.
\end{verbatim}
\normalsize
This rule says that at each time step $I \geq n1$, the agent must perform exactly one action. The constant, $n1$, is an integral part of this architecture: it represents the time step in which the new behavior mode is to take effect. For example, if we are in a scenario where we want the agent's initial behavior mode to be $b_0$ and switch to $b_1$ at time step $i$, then $n1=0$ for each time step $t<i$, and $n1=i$ for each time step $t\geq i$. This ensures that only the planned actions at time steps greater than or equal to $i$ have to obey the definition of behavior mode $b_1$. 

In each of the behavior mode's encodings, we additionally have several metrics that are calculated and considered by the agent when devising its plan, as in work by \cite{hi23}. What differentiates each of the behavior modes is the priority that is given to each of the metrics in the planning process, as described in Section \ref{sec:charlie}. For example, in the Safe behavior mode's encoding, we see the following ASP rule:
\small
\begin{verbatim}
    #maximize{ N4@4 : subgoal_count(N4);
               N3@3 : percentage_strongly_compliant(N3); 
               N2@2 : percentage_underspecified(N2);
               N1@1 : wait_count(N1)}.
\end{verbatim}
\normalsize
This says that the metric \verb|subgoal_count| should be prioritized first,  \verb|percentage_strongly_comp-| \verb|liant| should be prioritized second, \verb|percentage_underspecified| should be prioritized third, and \verb|wait_count| should be prioritized last. 
The \verb|subgoal_count| metric is a count of the number of subgoals that the agent completes during the plan. This is a novel inclusion in our proposed architecture. 
In the Mining Domain, the maximum number of subgoals that the agent can complete is three, one subgoal corresponding to the collection of each of the ores. The ASP encoding for this is:
\small
\begin{verbatim}
    subgoal(has_ore(gold)). subgoal(has_ore(silver)). subgoal(has_ore(iron)).
    subgoal_count(N) :- #count{F : subgoal(F), holds(F, n)} = N.
\end{verbatim}
\normalsize

The \verb|percentage_strongly_compliant| and \verb|percentage_underspecified| metrics come from work by Harders and Inclezan \cite{hi23}. 
Recall that a \textit{strongly-compliant} action is one that is explicitly permitted by the agent's policies and an \textit{underspecified} action is one that is neither permitted nor not permitted by the agent's policies. The safe agent prioritizes actions that are explicitly permitted, because it is designed to act in an extremely cautious way, even if unnecessary. Finally, the \verb|wait_count| metric is a count of the number of $wait$ actions in the agent's plan. The higher the \verb|wait_count|, the shorter the plan. Agents under this proposed architecture only perform waiting actions after they have completed as many subgoals as possible. This is encoded in ASP as:
\small
\begin{verbatim}
    :- occurs(wait, I1), occurs(A, I2), I2 > I1, I1 >= n1, A != wait.
\end{verbatim}
\normalsize

Now that we have an understanding of what each of these metrics represent, let's compare the prioritization order of the Safe agent to that of the Normal agent.
\small
\begin{verbatim}
    #maximize{ N4@4 : subgoal_count(N4);
               N3@3 : wait_count(N3);
               N2@2 : percentage_underspecified(N2); 
               N1@1 : percentage_strongly_compliant(N1)}.
\end{verbatim}
\normalsize
The Normal agent still prioritizes first completing as many subgoals as possible, but instead of also trying to maximize the number of strongly-compliant actions in the plan, it values a shorter plan. Additionally, both the Safe and Normal agent behavior modes have constraints saying that no non-compliant actions w.r.t. obligations are allowed.
The Risky agent only considers two metrics in its planning process, \verb|subgoal_count| and \verb|wait_count|, in that order. This allows the Risky agent to devise the shortest plan possible while completing as many subgoals as possible, with the trade-off that it completely disregards any policies that are imposed on it. The ASP encoding is:
\small
\begin{verbatim}
    #maximize{ N2@2: subgoal_count(N2); N1@1: wait_count(N1)}.
\end{verbatim}
\normalsize

Finally, the \textit{learned information} refers to the facts formed by \emph{holds} and \emph{occurs} literals that are true prior to the time step when the behavior mode modification took effect.

\subsection{Python Component}
The Python Component of the proposed architecture is what allows us to manage the behavior mode modification process. This component utilizes the \textsc{Clingo} Python API, which allows developers to solve ASP programs and analyze their output using Python code. For the Mining Domain, we present a class called \verb|MiningDomainSolver|, which takes as input a scenario number, 
an initial behavior mode, and a list of behavior mode changes and the time steps when they are to take effect. Once this class is instantiated, a user may call the class's function called \verb|generate_plan_with_bmode_changes()|, which returns the plan as a string. This function follows the control flow outlined below. It is also worth noting that this control flow is not specific to the Mining Domain, and can be applied to any other dynamic domain under this proposed architecture:
\begin{enumerate}
    \item $n1$ is computed. As mentioned previously, $n1 = 0$, when we are solving the ASP program corresponding to the initial behavior mode, and $n1$ is equal to the time step of each behavior mode change after that.
    \item The ASP program is created inside of a string variable by reading the contents of the text files of the dynamic domain (i.e., the dynamic domain encoding, scenario encoding, policy encoding, and behavior mode encoding). \emph{Learned information} (stored inside of a string variable) is also added to the ASP program. 
    \item A temporary text file is created and the ASP program is written to it.
    \item A \textsc{Clingo} \emph{control} object is created using the \textsc{Clingo} API, and the temporary file is loaded into it using its provided \verb|load()| function.
    \item The ASP program is solved using the \verb|solve()| function of the \emph{control} object. This function solves the loaded ASP program and outputs the literals in the answer set that are specified using \textsc{Clingo}'s \verb|#show| directive in the ASP program.
    \item If there is a behavior mode change, these literals are saved to a class variable and filtered to produce the \emph{learned information} for the next iteration.
\end{enumerate}

\section{Software System}
\label{sec:gui}
Next, let us discuss the graphical user interface (GUI) that was developed as a proof of concept for a program that allows a controller to make behavior mode modifications of an agent.
The software is available at \url{https://github.com/scglaze/MiningRobotDomainGUI}. 

\begin{minipage}{\textwidth}
  \begin{minipage}
  {0.51\textwidth}
    \centering
    \includegraphics[width=0.8\textwidth]
    {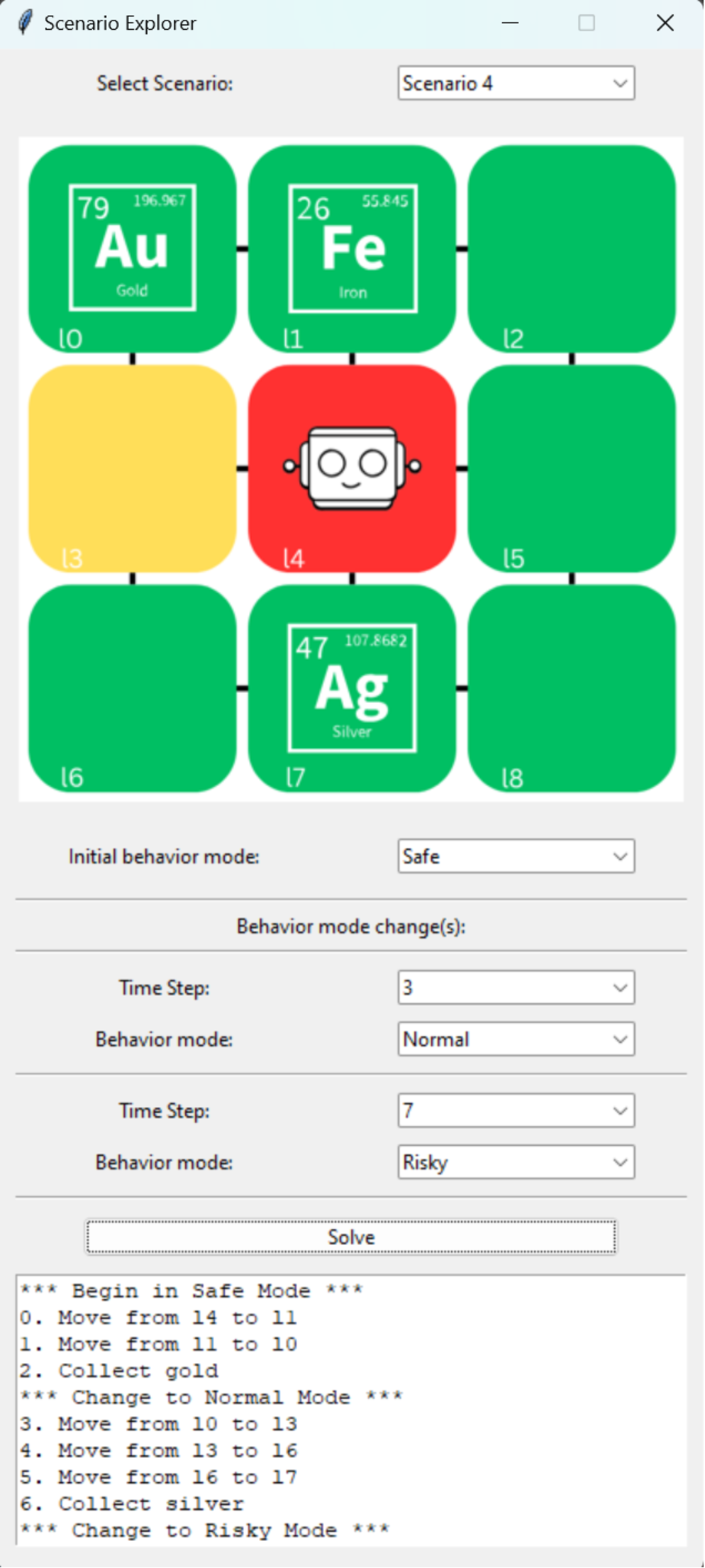}
    \label{fig:GUIScreenshot}
    \captionof{figure}{GUI screenshot with input parameters for the Mining Domain Scenario 4}
  \end{minipage}
  \hfill
  \begin{minipage}
  {0.47\textwidth}
    \centering
    \includegraphics[width=0.8\textwidth]
    {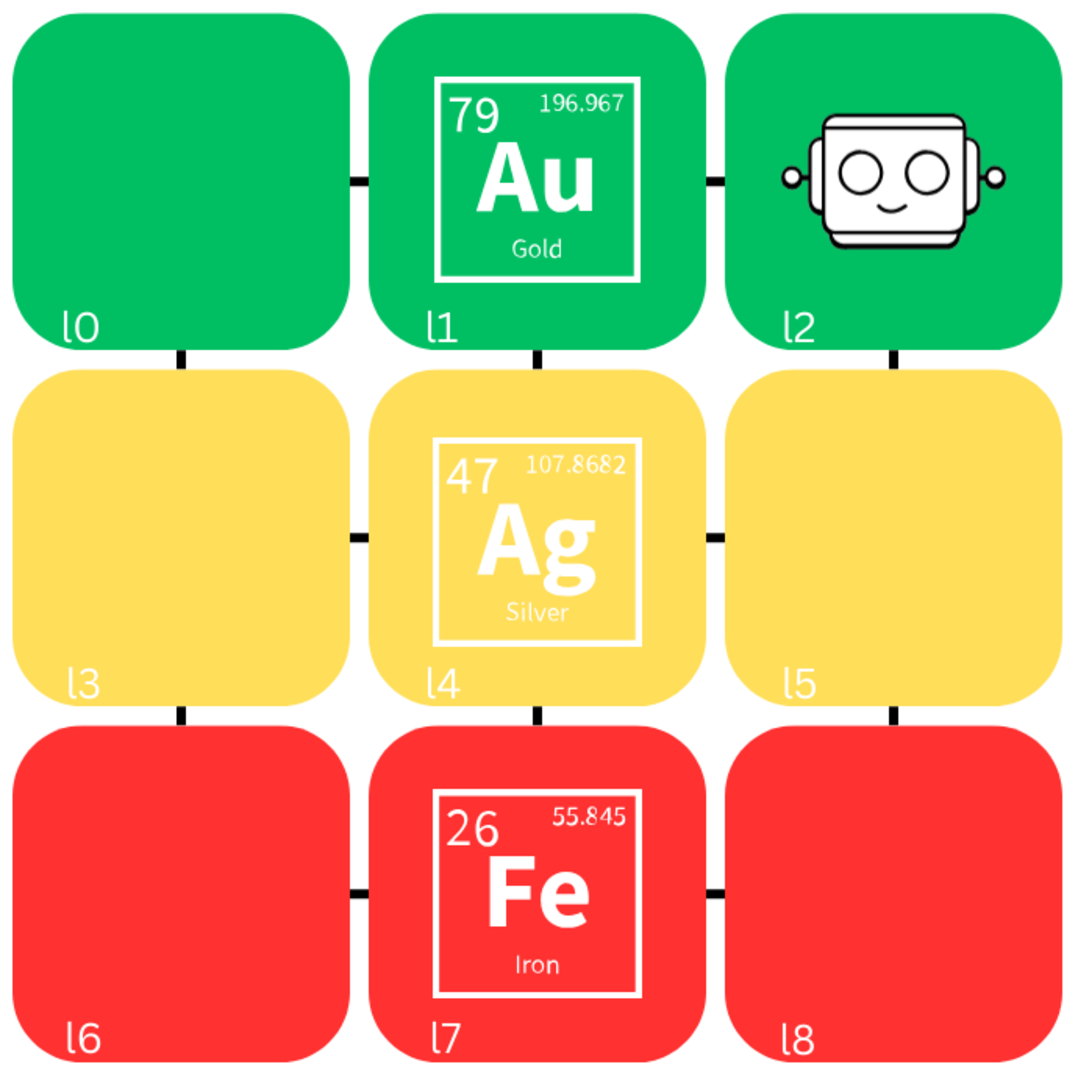}
    \label{fig:MiningRobotScenario9}
    \captionof{figure}{Mining Domain Scenario 9}
 \end{minipage}
\end{minipage}

We leveraged the Tkinter Python library for GUI development. The GUI allows a user to select one of the 10  scenarios that we have prepared from the Mining Domain, an initial behavior mode, and up to two behavior mode changes. When the user selects a scenario, a graphic for that scenario appears to serve as a visual aid. Once the user is finished inputting their desired parameters, there is a ``Solve" button that initiates the solving process. This solving process is performed by the aforementioned \verb|MiningDomainSolver| class described in the previous section, by feeding the user's input to it as its parameter. 
Before this is done, though, there are several validation checks that are performed. For example, the user must input both a behavior mode \emph{and} a time step for each behavior mode modification. If there are any validation checks that are violated, then a dialog box appears with a description of the error. If there are no validation errors, once the solving process is finished, the generated plan is displayed in a user-friendly manner in a text box at the bottom of the GUI, as shown in Figure 3.

\section{Evaluation}
\label{sec:evaluation}

\textbf{Runtime Performance:} 
We ran experiments on the 10 scenarios in the Mining Domain. We measured the runtime performance for each of the three behavior modes by themselves, 
and for the six combinations of first-order behavior mode modifications (i.e. only one modification made during the plan). We varied the time step when the modification occurred from scenario to scenario, based on what we subjectively deemed as leading to most illustrative changes in plans. Additionally, we measured the runtime performance of second-order behavior mode modifications for two of the scenarios that are more complex. 
We present the runtime performance of 
Scenario \#4 from Figure 3  in Table 3 
and Scenario \#9 from Figure 4 in Table 4. 
Time steps when behavior mode modifications are made are listed in parenthesis. Scenario \#9 involves second-order behavior mode modification. 
The runtime performance that we report does not come from the \textsc{Clingo} solver itself, but instead, is measured via code that was integrated into the Python component for this experiment. This test code utilizes the \verb|time| Python library. The reason we went this route instead of measuring the reported runtime from \textsc{Clingo}, is that our proposed architecture requires an additional \verb|solve()| for each behavior mode modification that is made. 
We ran each experiment 10 times, and report the mean in seconds (\textbf{T (s)}) and standard deviation (\textbf{SD}). All experiments were performed on a machine with an Intel(R) Core(TM) i7-1065G7 CPU @ 1.30GHz 1.50 GHz processor and 16 GB RAM. 


\medskip
\small
\begin{minipage}{\textwidth}
  \begin{minipage}[t]
  {0.4\textwidth}
    \centering
    \captionof{table}{Mining Domain Scenario 4: Runtime Data (Python + \textsc{Clingo})}
\label{tbl:Scenario4}
\begin{tabular}{lll}
\hline\hline
\textbf{Behavior Modes} & \textbf{T (s)} & \textbf{SD} \\ 
\hline
Safe & 1.684 & 0.12 \\ 
Normal & 1.379 & 0.08 \\ 
Risky & 0.181 & 0.08 \\  \hline
Safe to Normal (2) & 2.641 & 0.11 \\ 
Safe to Risky (2) & 1.881 & 0.06 \\ 
Normal to Safe (2) & 2.666 & 0.14 \\ 
Normal to Risky (2) & 1.764 & 0.24 \\ 
Risky to Safe (2) & 1.452 & 0.17 \\ 
Risky to Normal (2) & 0.944 & 0.07 \\ \hline\hline
\end{tabular}
  \end{minipage}
  \hfill
  \begin{minipage}[t]
  {0.58\textwidth}
    \centering
    \captionof{table}{Mining Domain Scenario 9: Runtime Data\\ (Python + \textsc{Clingo})}
\label{tbl:Scenario9}
\begin{tabular}{lll}
\hline\hline
\textbf{Behavior Modes} & \textbf{T (s)} & \textbf{SD} \\ \hline
Safe & 0.652 & 0.04 \\ 
Normal & 0.565 & 0.07 \\ 
Risky & 0.094 & 0.01 \\ \hline
Safe to Normal (2) & 1.059 & 0.08 \\ 
Safe to Risky (2) & 0.767 & 0.04 \\ 
Normal to Safe (2) & 0.963 & 0.12 \\ 
Normal to Risky (2) & 0.596 & 0.03 \\ 
Risky to Safe (2) & 0.516 & 0.03 \\ 
Risky to Normal (2) & 0.481 & 0.04 \\ 
Safe to Normal (2) to Risky (4) & 1.202 & 0.12 \\ 
Safe to Normal (3) to Risky (6) & 1.090 & 0.12 \\ \hline \hline
\end{tabular}
 \end{minipage}
\end{minipage}
\normalsize

\medskip
We see that the Safe behavior mode has a slightly longer runtime than that of the Normal behavior mode, and that the Risky behavior mode has the shortest runtime overall, which is a trend observed across all 10 scenarios. This is intuitive, as the Safe behavior mode takes the most amount of factors (i.e., aggregates) into consideration during plan generation, and the Risky behavior mode takes significantly less factors into consideration. Similarly, we observe that the runtime of behavior mode \textit{modifications} (with the time step when the modification occurs specified in parenthesis) follows this same trend -- modifications involving the Risky mode add less runtime than either of the other behavior modes, and the Safe mode adds the most to runtime. 
Scenario \#9 additionally tests second-order behavior mode modifications. We selected 
modifying the agent's behavior mode from Safe, to Normal, to Risky 
because the agent begins in a safe area of the grid, where the gold is also located. The silver is located adjacent to the gold but in a medium-risk location that the Safe agent cannot access. Therefore, the Safe agent will \emph{wait} indefinitely, unless there is a behavior mode modification. This behavior is mirrored by the Normal agent after it collects the silver. 
Hence, the plan generated by the agent with behavior mode parameters ``Safe to Normal (2) to Risky (4)" is set up to collect the ores without the Safe or Normal agents waiting at all, and the agent with behavior mode parameters ``Safe to Normal (3) to Risky (6)" is set up for the Safe and Normal agents to wait for exactly 1 time step before its behavior mode is modified to a more Risky one that allows them to complete another subgoal.
An interesting observation is that the agent with behavior mode parameters ``Safe to Normal (3) to Risky (6)" has a faster runtime than that with parameters ``Safe to Normal (2) to Risky (4)." We speculate that this is because the plan that is generated by the Normal agent at time step 3 has a shorter span of time steps to plan for than when starting at time step 2, and likewise for the Risky agent at time step 6 versus 4.

We also ran experiments on the 14 scenarios of the \textit{Room Domain} by Harders and Inclezan \cite{hi23}. The results are in Table 5 and they match the observations for the Mining Domain.

\small
\begin{table}[hbt!]
\caption{Performance Results: Room Domain (Python + \textsc{Clingo})}
\label{tbl:performance_room}
\centering
\begin{tabular*}{\linewidth}{@{\extracolsep{\fill}} c | c c | c c | c c || c c  l}
\hline\hline
                  & \multicolumn{2}{c | }{\textbf{Safe Mode}}  & \multicolumn{2}{c | }{\textbf{Normal Mode}}  & \multicolumn{2}{c || }{\textbf{Risky Mode}}  & \multicolumn{3}{c  }{\textbf{One Behavior Mode Change}} \\ \hline
\textbf{Scen. \#} & \textbf{T (s)}     & \textbf{SD} & \textbf{T (s)} & \textbf{SD} & \textbf{T (s)}     & \textbf{SD} & \textbf{T (s)} & \textbf{SD}  & \textbf{Change}\\ \hline
1        & 6.876&	0.38	&7.468&	0.21&	7.196&	0.20&	14.621&	0.34&	Safe to Normal (1)   \\ 
2        &   7.777&	0.50&	7.620&	0.56&	8.521&	2.40&	15.541	&2.31&	Risky to Safe (2)  \\ 
3        &   7.772&	0.14&	9.667&	2.95&	7.689&	0.48&	14.345&	0.21&	Safe to Risky (3)   \\ 
4        &   7.763&	0.21&	7.644&	0.24&	7.474&	0.55&	14.271&	0.37&	Risky to Safe (1)  \\ 
5        &   7.316&	0.15&	7.190&	0.06&	7.085&	0.09&	14.096&	0.23&	Risky to Normal (1)   \\ 
6        &   7.762&	0.40&	7.232&	0.17&	7.159&	0.18&	13.795&	0.19&	Safe to Normal (2)   \\ 
7        &   7.836&	0.38&	7.442&	0.31&	7.223&	0.18&	13.932&	0.48&	Risky to Normal (2)  \\ 
8        &   7.196	&0.09&	7.487&	0.10&	7.859&	0.68&	14.135&	0.24&	Safe to Risky (2)   \\ 
9        &  9.103	&0.12&	7.727&	0.18&	7.665&	0.20&	15.821&	0.39&	Safe to Risky (2)   \\ 
10        &  7.305&	0.15&	7.307&	0.17&	7.207&	0.11&	13.871&	0.36&	Normal to Safe (2)   \\ 
11       &  8.065&	0.41&	7.870&	0.21&	7.706&	0.22&	14.882&	0.27&	Normal to Risky (1)   \\ 
12       &  7.741&	0.17&	7.689&	0.42&	7.488&	0.09&	14.520&	0.24&	Normal to Safe (2)   \\ 
13       &   7.762&	0.30&	7.712&	0.21&	7.739&	0.35&	14.720&	0.13&	Safe to Normal (2)  \\ 
14       &  8.369&	0.27&	7.395&	0.16&	7.334&	0.11&	13.346&	0.44&	Normal to Safe (4)  \\ 
        \hline \hline
\end{tabular*}
\end{table}
\normalsize

\noindent
\textbf{GUI Usability Study:} 
The final evaluation was a usability study for the GUI that we presented in Section 5. 
Our participants ($N = 6$) 
were given a brief explanation of the Mining Domain, and necessary background information on ASP planning. Then, they were asked to download and run an executable file for the GUI seen in Figure 3, and  
to test all 10 scenarios with different behavior mode parameters. Finally, they answered questions on a 5-point Likert scale. The average score and standard deviation for each question's responses are reported in Table~\ref{tbl:usability_study}.
While scores were generally high, especially for question 6, we do note the lower scores for questions 1, 2, 8. This indicates that the prototype GUI can be improved by using more modern-looking widgets, facilitating the process of downloading it, and providing more descriptive error messages that are displayed when input validation checks that are violated.

\begin{table}[bht]
\caption{Usability Study Results}
\label{tbl:usability_study}
\begin{minipage}{\textwidth}
\centering
\begin{tabular}{l l  c c}
\hline\hline
  & \textbf{Survey Question}                                                                                                    & \textbf{Average Score}  & \textbf{SD}   \\
  & & (scale 1-5) & \\
  \hline
1 & The executable (.exe) file was easy to download and run.                                                    & 3.83 & 1.47 \\
2 & The GUI has a nice look and feel.                                                                           & 3.60 & 0.89 \\
3 & The GUI was easy to interact with.                                                                          & 4.67 & 0.52 \\
4 & I did not encounter any odd behavior from the GUI.                                                          & 4.83 & 0.41 \\
5 & The images depicting the different scenarios were a useful & 4.50 & 0.55 \\
  & resource for   understanding the generated plan. &  &  \\
6 & It was easy to change behavior modes.                                                                       & 5.00 & 0.00 \\
7 & I understand the plan that was generated by the program.                                                    & 4.50 & 0.84 \\
8 & Error messages were easy to understand (Only answer   & 3.67 & 1.15\\
 & this question if you   received error messages).   &  & \\
 \hline\hline
\end{tabular}
\end{minipage}
\end{table}

\section{Related Work}
\label{sec:related}

Our work expands on Harders and Inclezan's \cite{hi23} notions of behavior modes w.r.t. norm-compliance. Another work on norm-aware agents is that by by Meyer and Inclezan \cite{mi21} who created the \apia architecture for norm-aware {\em intentional} agents. \apia agents operate with {\em activities} instead of simple plans, by building upon the AIA architecture by \cite{BlountGB15}. 
\apia agents can reason about agent intentions, 
but does not allow the agent's controller to easily set and change behavior modes. 
\cite{ShamsVPV17} introduced an ASP framework for reasoning and planning with norms for autonomous agents. The agent actions in their framework have an associated duration and can incur penalties, while policies have an expiration deadline. On the other hand, their framework does not model different behavior modes and changes between behavior modes, which is the focus of this paper. Other existing approaches on norm-aware agents focus solely on compliant behavior (e.g., \cite{Oren11,Alechina12}), while we were interested in studying a range of behavior modes on a spectrum for norm-abiding to non-compliant to enable the simulation of human behavior as well.
In our work, we assume that changes between behavior modes are justified in certain situations, such as emergency rescue operations, and this should be modeled and simulated. The question of emergency situations in relation to norms was previous studied by Alves and Fernández \cite{af17}, but only in the context of access control policies. In contrast, the use \aopl for norm specification in our architecture allows us to express not only access control policies (i.e., authorizations), but also obligations, both strict and defeasible, and preferences between policy statements.
In terms of defining behavior modes via priorities between different metrics, our work indicates some connections to Son and Pontelli's $\mathscr{PP}$ for specifying basic preferences \cite{son_pontelli_2006}. It is not clear though whether maximizations of percentage metrics, which occur in our description of behavior modes, can be achieved within the $\mathscr{PP}$ framework.



\section{Conclusions and Future Work}
\label{sec:conclusions}

We presented an ASP framework that defines how the controller of norm-aware autonomous agents can modify their behavior modes under the plan-choosing framework proposed by \cite{hi23}. We introduced a Python component that includes a wrapper class that can be used as the behavior mode-changing mechanism and a GUI with the potential for generalization, for other domains as well. 

In the future, one could generalize this proposed ASP simulation framework so that a controller can manipulate multiple agents' behavior modes as they work toward achieving their goal(s). Optimizing ASP encodings and Python code is another future goal, as it would allow for larger and more complicated dynamic domains to be simulated.
Finally, one could continue to develop additional behavior modes, outside of the three that are used in this framework. This would allow for more nuanced agent behavior to be modeled under this framework and generally in ASP.

\bibliographystyle{eptcs}
\bibliography{iclp2024}

\begin{thebibliography}{10}
\providecommand{\bibitemdeclare}[2]{}
\providecommand{\surnamestart}{}
\providecommand{\surnameend}{}
\providecommand{\urlprefix}{Available at }
\providecommand{\url}[1]{\texttt{#1}}
\providecommand{\href}[2]{\texttt{#2}}
\providecommand{\urlalt}[2]{\href{#1}{#2}}
\providecommand{\doi}[1]{doi:\urlalt{https://doi.org/#1}{#1}}
\providecommand{\eprint}[1]{arXiv:\urlalt{https://arxiv.org/abs/#1}{#1}}
\providecommand{\bibinfo}[2]{#2}

\bibitemdeclare{inproceedings}{Alechina12}
\bibitem{Alechina12}
\bibinfo{author}{Natasha \surnamestart Alechina\surnameend},
  \bibinfo{author}{Mehdi \surnamestart Dastani\surnameend} \&
  \bibinfo{author}{Brian \surnamestart Logan\surnameend}
  (\bibinfo{year}{2012}): \emph{\bibinfo{title}{Programming norm-aware
  agents}}.
\newblock In: {\slshape \bibinfo{booktitle}{Proceedings of the 11th
  International Conference on Autonomous Agents and Multiagent Systems - Volume
  2}}, \bibinfo{series}{AAMAS '12}, \bibinfo{publisher}{International
  Foundation for Autonomous Agents and Multiagent Systems},
  \bibinfo{address}{Richland, SC}, p. \bibinfo{pages}{1057–1064},
  \doi{10.5555/2343776.2343848}.

\bibitemdeclare{article}{af17}
\bibitem{af17}
\bibinfo{author}{Sandra \surnamestart Alves\surnameend} \&
  \bibinfo{author}{Maribel \surnamestart Fernandez\surnameend}
  (\bibinfo{year}{2017}): \emph{\bibinfo{title}{A graph-based framework for the
  analysis of access control policies}}.
\newblock {\slshape \bibinfo{journal}{Theoretical Computer Science}}
  \bibinfo{volume}{685}, pp. \bibinfo{pages}{3--22},
  \doi{10.1016/j.tcs.2016.10.018}.

\bibitemdeclare{inproceedings}{BlountGB15}
\bibitem{BlountGB15}
\bibinfo{author}{Justin \surnamestart Blount\surnameend},
  \bibinfo{author}{Michael \surnamestart Gelfond\surnameend} \&
  \bibinfo{author}{Marcello \surnamestart Balduccini\surnameend}
  (\bibinfo{year}{2015}): \emph{\bibinfo{title}{A Theory of Intentions for
  Intelligent Agents - ({E}xtended {A}bstract)}}.
\newblock In: {\slshape \bibinfo{booktitle}{Proceedings of the 13th
  International Conference on Logic Programming and Nonmonotonic Reasoning}},
  {\slshape \bibinfo{series}{LNCS}} \bibinfo{volume}{9345},
  \bibinfo{publisher}{Springer}, pp. \bibinfo{pages}{134--142},
  \doi{10.1007/978-3-319-23264-5\_12}.

\bibitemdeclare{inproceedings}{CorapiVPRS10}
\bibitem{CorapiVPRS10}
\bibinfo{author}{Domenico \surnamestart Corapi\surnameend},
  \bibinfo{author}{Marina~De \surnamestart Vos\surnameend},
  \bibinfo{author}{Julian~A. \surnamestart Padget\surnameend},
  \bibinfo{author}{Alessandra \surnamestart Russo\surnameend} \&
  \bibinfo{author}{Ken \surnamestart Satoh\surnameend} (\bibinfo{year}{2010}):
  \emph{\bibinfo{title}{Norm Refinement and Design through Inductive
  Learning}}.
\newblock In \bibinfo{editor}{Marina~De \surnamestart Vos\surnameend},
  \bibinfo{editor}{Nicoletta \surnamestart Fornara\surnameend},
  \bibinfo{editor}{Jeremy~V. \surnamestart Pitt\surnameend} \&
  \bibinfo{editor}{George~A. \surnamestart Vouros\surnameend}, editors:
  {\slshape \bibinfo{booktitle}{Coordination, Organizations, Institutions, and
  Norms in Agent Systems {VI} - {COIN} 2010 International Workshops}},
  {\slshape \bibinfo{series}{Lecture Notes in Computer Science}}
  \bibinfo{volume}{6541}, \bibinfo{publisher}{Springer}, pp.
  \bibinfo{pages}{77--94}, \doi{10.1007/978-3-642-21268-0\_5}.

\bibitemdeclare{article}{gi13}
\bibitem{gi13}
\bibinfo{author}{Michael \surnamestart Gelfond\surnameend} \&
  \bibinfo{author}{Daniela \surnamestart Inclezan\surnameend}
  (\bibinfo{year}{2013}): \emph{\bibinfo{title}{Some properties of system
  descriptions of {AL}d}}.
\newblock {\slshape \bibinfo{journal}{J. Appl. Non Class. Logics}}
  \bibinfo{volume}{23}(\bibinfo{number}{1-2}), pp. \bibinfo{pages}{105--120},
  \doi{10.1080/11663081.2013.798954}.

\bibitemdeclare{book}{gk14}
\bibitem{gk14}
\bibinfo{author}{Michael \surnamestart Gelfond\surnameend} \&
  \bibinfo{author}{Yulia \surnamestart Kahl\surnameend} (\bibinfo{year}{2014}):
  \emph{\bibinfo{title}{Knowledge Representation, Reasoning, and the Design of
  Intelligent Agents}}.
\newblock \bibinfo{publisher}{Cambridge University Press},
  \doi{10.1017/CBO9781139342124}.

\bibitemdeclare{article}{gl91}
\bibitem{gl91}
\bibinfo{author}{Michael \surnamestart Gelfond\surnameend} \&
  \bibinfo{author}{Vladimir \surnamestart Lifschitz\surnameend}
  (\bibinfo{year}{1991}): \emph{\bibinfo{title}{{C}lassical {N}egation in
  {L}ogic {P}rograms and {D}isjunctive {D}atabases}}.
\newblock {\slshape \bibinfo{journal}{New Generation Computing}}
  \bibinfo{volume}{9}(\bibinfo{number}{3/4}), pp. \bibinfo{pages}{365--386},
  \doi{10.1007/BF03037169}.

\bibitemdeclare{inproceedings}{gl08}
\bibitem{gl08}
\bibinfo{author}{Michael \surnamestart Gelfond\surnameend} \&
  \bibinfo{author}{Jorge \surnamestart Lobo\surnameend} (\bibinfo{year}{2008}):
  \emph{\bibinfo{title}{Authorization and {Obligation} {Policies} in {Dynamic}
  {Systems}}}.
\newblock In \bibinfo{editor}{Maria \surnamestart Garcia de~la
  Banda\surnameend} \& \bibinfo{editor}{Enrico \surnamestart
  Pontelli\surnameend}, editors: {\slshape \bibinfo{booktitle}{Logic
  {Programming}}}, \bibinfo{series}{Lecture {Notes} in {Computer} {Science}},
  \bibinfo{publisher}{Springer}, \bibinfo{address}{Berlin, Heidelberg}, pp.
  \bibinfo{pages}{22--36}, \doi{10.1007/978-3-540-89982-2\_7}.

\bibitemdeclare{inproceedings}{hi23}
\bibitem{hi23}
\bibinfo{author}{Charles \surnamestart Harders\surnameend} \&
  \bibinfo{author}{Daniela \surnamestart Inclezan\surnameend}
  (\bibinfo{year}{2023}): \emph{\bibinfo{title}{Plan Selection Framework for
  Policy-Aware Autonomous Agents}}.
\newblock In \bibinfo{editor}{Sarah~Alice \surnamestart Gaggl\surnameend},
  \bibinfo{editor}{Maria~Vanina \surnamestart Martinez\surnameend} \&
  \bibinfo{editor}{Magdalena \surnamestart Ortiz\surnameend}, editors:
  {\slshape \bibinfo{booktitle}{Logics in Artificial Intelligence - 18th
  European Conference, {JELIA}}}, {\slshape \bibinfo{series}{LNCS}}
  \bibinfo{volume}{14281}, \bibinfo{publisher}{Springer}, pp.
  \bibinfo{pages}{638--646}, \doi{10.1007/978-3-031-43619-2\_43}.

\bibitemdeclare{article}{di23}
\bibitem{di23}
\bibinfo{author}{Daniela \surnamestart Inclezan\surnameend}
  (\bibinfo{year}{2023}): \emph{\bibinfo{title}{An {ASP} Framework for the
  Refinement of Authorization and Obligation Policies}}.
\newblock {\slshape \bibinfo{journal}{Theory and Practice of Logic
  Programming}} \bibinfo{volume}{23}(\bibinfo{number}{4}), p.
  \bibinfo{pages}{832–847}, \doi{10.1017/S147106842300011X}.

\bibitemdeclare{incollection}{mt99}
\bibitem{mt99}
\bibinfo{author}{Victor~W. \surnamestart Marek\surnameend} \&
  \bibinfo{author}{Miroslaw \surnamestart Truszczynski\surnameend}
  (\bibinfo{year}{1999}): \emph{\bibinfo{title}{Stable Models and an
  Alternative Logic Programming Paradigm}}.
\newblock In \bibinfo{editor}{Krzysztof~R. \surnamestart Apt\surnameend},
  \bibinfo{editor}{Victor~W. \surnamestart Marek\surnameend},
  \bibinfo{editor}{Mirek \surnamestart Truszczynski\surnameend} \&
  \bibinfo{editor}{David~Scott \surnamestart Warren\surnameend}, editors:
  {\slshape \bibinfo{booktitle}{The Logic Programming Paradigm - {A} 25-Year
  Perspective}}, \bibinfo{series}{Artificial Intelligence},
  \bibinfo{publisher}{Springer}, pp. \bibinfo{pages}{375--398},
  \doi{10.1007/978-3-642-60085-2\_17}.

\bibitemdeclare{inproceedings}{mi21}
\bibitem{mi21}
\bibinfo{author}{John \surnamestart Meyer\surnameend} \&
  \bibinfo{author}{Daniela \surnamestart Inclezan\surnameend}
  (\bibinfo{year}{2021}): \emph{\bibinfo{title}{{APIA:} An Architecture for
  Policy-Aware Intentional Agents}}.
\newblock In: {\slshape \bibinfo{booktitle}{Proceedings of the 37th
  International Conference on Logic Programming (Technical Communications)}},
  {\slshape \bibinfo{series}{{EPTCS}}} \bibinfo{volume}{345}, pp.
  \bibinfo{pages}{84--98}, \doi{10.4204/EPTCS.345.23}.

\bibitemdeclare{inproceedings}{Oren11}
\bibitem{Oren11}
\bibinfo{author}{Nir \surnamestart Oren\surnameend}, \bibinfo{author}{Wamberto
  \surnamestart Vasconcelos\surnameend}, \bibinfo{author}{Felipe \surnamestart
  Meneguzzi\surnameend} \& \bibinfo{author}{Michael \surnamestart
  Luck\surnameend} (\bibinfo{year}{2011}): \emph{\bibinfo{title}{Acting on Norm
  Constrained Plans}}.
\newblock In \bibinfo{editor}{Jo{\~a}o \surnamestart Leite\surnameend},
  \bibinfo{editor}{Paolo \surnamestart Torroni\surnameend},
  \bibinfo{editor}{Thomas \surnamestart {\AA}gotnes\surnameend},
  \bibinfo{editor}{Guido \surnamestart Boella\surnameend} \&
  \bibinfo{editor}{Leon \surnamestart van~der Torre\surnameend}, editors:
  {\slshape \bibinfo{booktitle}{Computational Logic in Multi-Agent Systems}},
  \bibinfo{publisher}{Springer Berlin Heidelberg}, \bibinfo{address}{Berlin,
  Heidelberg}, pp. \bibinfo{pages}{347--363},
  \doi{10.1007/978-3-642-22359-4_24}.

\bibitemdeclare{article}{ShamsVPV17}
\bibitem{ShamsVPV17}
\bibinfo{author}{Zohreh \surnamestart Shams\surnameend},
  \bibinfo{author}{Marina~De \surnamestart Vos\surnameend},
  \bibinfo{author}{Julian~A. \surnamestart Padget\surnameend} \&
  \bibinfo{author}{Wamberto~Weber \surnamestart Vasconcelos\surnameend}
  (\bibinfo{year}{2017}): \emph{\bibinfo{title}{Practical reasoning with norms
  for autonomous software agents}}.
\newblock {\slshape \bibinfo{journal}{Eng. Appl. Artif. Intell.}}
  \bibinfo{volume}{65}, pp. \bibinfo{pages}{388--399},
  \doi{10.1016/J.ENGAPPAI.2017.07.021}.

\bibitemdeclare{article}{son_pontelli_2006}
\bibitem{son_pontelli_2006}
\bibinfo{author}{Tran~Cao \surnamestart Son\surnameend} \&
  \bibinfo{author}{Enrico \surnamestart Pontelli\surnameend}
  (\bibinfo{year}{2006}): \emph{\bibinfo{title}{Planning with preferences using
  logic programming}}.
\newblock {\slshape \bibinfo{journal}{Theory and Practice of Logic
  Programming}} \bibinfo{volume}{6}(\bibinfo{number}{5}), p.
  \bibinfo{pages}{559–607}, \doi{10.1017/S1471068406002717}.

\end{thebibliography}
\end{document}